\documentclass[twocolumn,showpacs,preprintnumbers,
amsmath,amssymb,aps,prl]{revtex4}
\def\be{\begin{equation}}
\def\ee{\end{equation}}
\def\bea{\begin{eqnarray}}
\def\eea{\end{eqnarray}}
\def\nn{\nonumber}
\def\p{\partial}
\def\half{\frac{1}{2}}
\def\mA{\mathcal{A}}

\begin{document}
\preprint{USTC-ICTS-04-10}
\preprint{gr-qc/0405029}
\title{New Formulation of the First Law of Black Hole Thermodynamics: A Stringy Analogy}
\author{Shuang-Qing Wu$^{*1,2}$}
\affiliation{\centerline{$^1$Department of Physics, Central China Normal University,
Wuhan, Hubei 430079, People's Republic of China} \\
\centerline{$^2$Interdisciplinary Center for Theoretical Study,
University of Science and Technology of China,} \\
\centerline{Hefei, Anhui 230026, People's Republic of China}}
\date{6 May 2004}
\revised{26 July 2004}

\begin{abstract}
We consider the first laws of thermodynamics for a pair of systems made up of the two horizons of a
Kerr-Newman black hole. These two systems are constructed in such a way that we only demand their
``horizon areas'' to be the sum and difference of that of the outer and inner horizons of their
prototype. Remarkably, these two copies bear a striking resemblance to the right- and left-movers
in string theory and D-brane physics. Our reformulation of the first law of black hole thermodynamics
can be thought of as an analogy of thermodynamics of effective string or D-brane models.
\end{abstract}

\maketitle

It is well-known that black hole thermodynamics was established in the late 1960s and 1970s through a
close analogy between the ordinary laws of thermodynamics and certain laws of black hole mechanics \cite{BCH}.
In particular, the first law of black hole thermodynamics is formulated as a relationship between the first
order variation of the horizon area and those of global charges \cite{GC} such as the mass, angular momentum,
and electric charge. This law makes it possible to assign an entropy to the horizon area and a temperature
to the surface gravity.

Black hole statistical mechanics requires to provide a reasonable microscopic description to the macroscopic
thermodynamics. The microscopic statistical derivation of black hole entropy plays a crucial role in our
understanding the puzzle of its origin problem \cite{BHEO,SV}. In the context of dual conformal field theory
\cite{CFT}, Carlip \cite{SC} argued that black hole thermodynamics and the statistical origin of entropy are
controlled universally by the near-horizon conformal symmetry. This symmetry should play an important role in
explaining the black hole phenomena. Many evidences from studies of dynamic processes such as Hawking radiation
\cite{SNS}, greybody factors \cite{GBF}, and more recently quasinormal modes as well as non-quasinormal modes
\cite{BC} strongly suggest that Carlip's method may take effect on the subject.

Recent developments in string theory (especially D-brane physics) have provided another fruitful framework to
consider quantum properties of black holes. The D-brane technology \cite{SV} counts the microstates for the
entropy by representing black holes as bound states of D-branes that carry the same energy and charges as the
black hole. It is also possible to apply the black hole/string correspondence principle \cite{SHP} to understand
the microscopic origin of black hole entropy. In the spirt of this principle, black holes can be modelled by
effective or fundamental strings \cite{CL} or be described microscopically as quantum states in string theory.
In this way, the black hole entropy is attributed to the sum of contributions from right- and left-moving
excitations of the string states.

A characteristic feature of effective strings or D-branes is that their spectrum divides into two distinct
sectors associated with the right (R) and left (L) moving excitations, respectively. The entropy of a system
consisting of these two weakly-coupled R- and L-moving modes is the sum of two contributions: $S = S_R +S_L$.
The temperature of this combined system is the harmonic mean of that of its parts:
\be\label{temp}
 \frac{2}{T_H} = \frac{1}{T_R} +\frac{1}{T_L} \, ,
\ee
and it is usually identified with the Hawking temperature of a black hole.

The temperature formula (\ref{temp}) appears frequently in the studies of black hole dynamic phenomena such as
absorption cross section, emission rates, decay rates, greybody factors \cite{GBF}, and moreover quasinormal
and non-quasinormal modes \cite{BC}. It is our curiosity to understand why its appearance is so universal in
all the cases. In what follows, we present a thermodynamic framework to appreciate it. To do so, we take a new
look at the classical thermodynamics of a Kerr-Newman (KN) black hole \cite{KN} and wish to seek some clues
on relating it with string theory, especially with the effective string/D-brane picture of black holes in a
pure thermodynamic side. Specifically, we consider the first law thermodynamics of a pair of systems constructed
from the two horizons of a KN black hole. It is interesting to observe that these two systems bear a striking
resemblance to the right- and left-movers in string theory and D-brane physics. We point out that there exists
a close relationship between the thermodynamics of these systems and that of effective string or D-brane models.
A remarkable advantage of our discussion presented here is that it relies on the validity in that the laws of
black hole mechanics does not depend upon the details of the underlying dynamical theory, even without knowledge
of fundamental string theory.

To begin with, recall that the usual thermodynamics of a four-dimensional charged rotating (KN) black hole
with three classical parameters: the mass $M$, charge $Q$, and angular momentum $J = Ma$. For our discussion
here, we only focus on the first law of black hole thermodynamics in this most general case. It is essentially
formulated as the Bekenstein-Smarr (B-S) differential mass formula \cite{BS}:
\be
 dM = \half\kappa~d\mA +\Omega~dJ + \Phi~dQ \, ,
\ee
where $\kappa$, $\Omega$, and $\Phi$ denote, respectively, the surface gravity, angular velocity, and electrostatic
potential of the event horizon of a KN black hole. (Here and hereafter, units are adopted where $G = c = \hbar = k_B
= 1$, and it is more economical to use the notation of a ``reduced'' horizon area $\mA = A/(4\pi)$ rather than the
horizon area $A$, so that the Bekenstein-Hawking entropy is $S = A/4 = \pi\mA$.)

The KN black hole has two horizons: the inner horizon (Cauchy horizon) $r_- = M -\epsilon$ and the outer horizon
(event horizon) $r_+ = M +\epsilon$, where $\epsilon = \sqrt{M^2 -Q^2 -a^2}$. In particular, the thermodynamics
associated with the outer event horizon of the black hole is related to the fundamental process of Hawking
radiation. Similarly, one can prove that the thermodynamics of the inner Cauchy horizon is associated with
another quantum process of Hawking ``absorption'' \cite{ZZL,WC}. Namely, for an observer rest at the infinity
he observes a net flux of Hawking radiation outgoing from the event horizon to the infinity, while for an
imaginary observer inhabiting in the intrinsic singular region (inside the Cauchy horizon), he will observe
a flux of Hawking ``absorption'' ingoing from the intrinsic singular region to the inner horizon also. Because
a KN black hole is stationary, its outer horizon is in thermal equilibrium with the thermal radiation outside
the black hole. We think that its inner horizon is certainly in thermal equilibrium with the thermal radiation
in the intrinsic singular region \cite{ZZL}.

As is well-known, the outer horizon of a KN black hole is treated as a single thermodynamical system, and the
conventional four laws of thermodynamics are just stated for this system. Similarly, if one treats the inner
horizon as an independent thermodynamical system, one can establish another four laws of thermodynamics for
the inner horizon also \cite{WC}. The well-known B-S integral and differential mass formulae for the first
laws of thermodynamics corresponding to the inner and outer horizons of a general, non-extremal KN black hole
are given by \cite{WC,ZZL,CF}
\bea\label{FLio}
  M &=& ~~~~\kappa_{\pm}\mA_{\pm} +2\Omega_{\pm}J +\Phi_{\pm}Q \, , \nn \\
 dM &=& \half\kappa_{\pm}d\mA_{\pm} +\Omega_{\pm}dJ +\Phi_{\pm}dQ \, ,
\eea
where the ``reduced'' horizon area $\mA_{\pm} = r^2_{\pm} +a^2 = 2Mr_{\pm} -Q^2$, while the surface gravity
$\kappa_{\pm}$, angular velocity $\Omega_{\pm}$, and electric potential $\Phi_{\pm}$ at the two horizons are
\cite{WC}
\be\label{KNQs}
 \kappa_{\pm} = \frac{r_{\pm} -M}{\mA_{\pm}} = \frac{\pm\epsilon}{\mA_{\pm}} \, , ~~
 \Omega_{\pm} = \frac{a}{\mA_{\pm}} \, , ~~ \Phi_{\pm} = \frac{Qr_{\pm}}{\mA_{\pm}} \, ,
\ee
respectively. In general, the B-S mass formulae can be derived from Christodoulou mass-squared formula \cite{CR}
for the reversible and irreversible processes by using thermodynamical relations. Specifically to a general,
nonextremal KN black hole, the Christodoulou formula \cite{CR} reads
\be\label{CR}
 M^2 = \frac{\mA_+ +\mA_-}{4} +\frac{Q^2}{2} = \frac{\mA_{\pm}}{4} +\frac{4J^2+Q^4}{4\mA_{\pm}} +\frac{Q^2}{2} \, .
\ee
From the identity (\ref{CR}), one can derive the expressions for the three thermodynamical quantities in Eq.
(\ref{KNQs}) by the following thermodynamical relations
\be
 \kappa_{\pm} = 2\frac{\p M}{\p\mA_{\pm}}\Big|_{J,Q}\, , ~
 \Omega_{\pm} = \frac{\p M}{\p J}\Big|_{Q,\mA_{\pm}}\, , ~
 \Phi_{\pm} = \frac{\p M}{\p Q}\Big|_{J,\mA_{\pm}}\, .
\ee

As a KN black hole is a system of two horizons, one should consider both of them for the sake of completeness.
Considerable research \cite{CL,ZZL,WC,CF} has confirmed this point. For instance, the importance of the inner
horizon is also emphasized in the viewpoint of string theory \cite{CL}. Just as did in Ref. \cite{ZZL}, we may
think a KN black hole can be regarded as a composite thermodynamical system composed of two independent subsystems,
its outer and inner horizons \cite{ZZL}. For each subsystem, the entropy of each horizon is proportional to
its horizon area, and the temperature is proportional to its surface gravity, namely, $S_{\pm} = \pi \mA_{\pm}$,
and $T_{\pm} = \kappa_{\pm}/(2\pi)$.

Now we present our main idea. Using the original inner and outer horizons as blocks, we can construct two new
thermodynamical systems called respectively as the R-system and L-system by demanding their ``reduced'' areas
to be the sum and difference of that of the outer and inner horizons of a KN black hole
\bea
 \mA_R &=& \mA_+ + \mA_- = 4M^2 -2Q^2 \, , \nn \\
 \mA_L &=& \mA_+ -\mA_- = 4M\epsilon \, . \quad
\eea
This idea is indeed supported by researches done in Ref. \cite{CL} from the viewpoint of string theory, however
it is reverse to the observation of Cvetic and Larsen [11] who find that the left- and right-moving thermodynamics
of the string theory correspond to the sum and the difference of the outer and the inner horizon thermodynamics.

The purpose of this Letter is to see whether or not we can understand the first law of thermodynamics of the
original KN black hole through exploring the thermodynamic features of these two new systems. In one way we
identify each system with a black hole. As the original motherboard carries three classical hairs ($M, Q, J$),
by the definition of our systems, it is immediately recognized that the R-system carries only two hairs ($M, Q$),
and the L-system carries three hairs ($M, Q, J$). So they are two distinct copies describing two different kinds
of black holes. In particular, there exists an asymmetry of angular momentum between the R- and L-systems as we
will see below. The disappearance of angular momentum in the R-system and its presence in the L-system imply
that the black hole degrees of freedom relevant for near-BPS excitations can be described by a $(0, 4)$ chiral
superconformal field theory with an SU(2) current algebra associated to rotations \cite{GBF}.

It is now a position to reveal the thermodynamic properties of our new systems. To investigate their thermodynamic
features in detail, let us focus firstly on the R-system. For this system, the Christodoulou-type mass-squared
formula is simply written as
\be\label{CR1}
 M^2 = \frac{\mA_R}{4} +\frac{Q^2}{2} \, .
\ee
From this relation, it is easily to deduce the integral and differential B-S mass formula as follows:
\bea\label{FL1}
  M &=& \frac{\mA_R}{4M} +\frac{Q^2}{2M} ~= ~~~~\kappa_R\mA_R +\Phi_R Q \, , \nn \\
 dM &=& \frac{d\mA_R}{8M} +\frac{Q^2}{2M} = \half\kappa_R d\mA_R +\Phi_R dQ \, ,
\eea
where the surface gravity $\kappa_R = 1/(4M)$, and electric potential $\Phi_R = Q/(2M)$ can also be obtained
by the following thermodynamical definitions
\bea
 && \kappa_R  = 2\frac{\p M}{\p\mA_R}\Big|_Q = \frac{1}{4M} = \frac{\epsilon}{\mA_L} \, , \nn \\
 && \Phi_R  = \frac{\p M}{\p Q}\Big|_{\mA_R} = \frac{Q}{2M} = \frac{2Q\epsilon}{\mA_L} \, .
\eea

The R-system can be viewed as describing a nonrotating charged black hole with mass $M$ and charge $Q$, while
its entropy and temperature being given by
\bea\label{ent1}
 S_R &=& \pi\mA_R = \pi(\mA_+ +\mA_-) = S_+ +S_- \nn \\
     &=& 2\pi(2M^2 -Q^2) \, , \nn \\
 T_R &=& \frac{\kappa_R}{2\pi} = \frac{1}{8\pi M} \, .
\eea
It is not difficult to recognize that this system is thermodynamically equivalent to the GHS dilatonic black
hole \cite{GHS} with the dilaton parameter $\phi_0 = 0$, whose horizon is at $r_h = 2M$, its surface gravity
is $\kappa_h = 1/(4M)$, and ``reduced'' horizon area is just $\mA_h = r_h(r_h -Q^2/M) = \mA_R$.

Now we turn to the L-system. The above derivation is also applicable to this case, but with a little involved
algebras. The corresponding Christodoulou-type formula is given by
\be\label{CR2}
 M^2 = \frac{Q^2 +\sqrt{Q^4 +4J^2 +\mA_L^2/4}}{2} \, ,
\ee
and the first law is formulated as follows
\bea\label{FL2}
  M &=& ~~~~\kappa_L\mA_L +2\Omega_L J +\Phi_L Q \, , \nn \\
 dM &=& \half\kappa_L d\mA_L +\Omega_L dJ +\Phi_L dQ \, ,
\eea
where the surface gravity $\kappa_L$, angular velocity $\Omega_L$, and electric potential $\Phi_L$ for this
system are determined by the following thermodynamical expressions
\bea
 && \kappa_L = 2\frac{\p M}{\p\mA_L}\Big|_{J,Q} = \frac{\epsilon}{\mA_R}
  = \frac{\sqrt{M^2 -Q^2 -a^2}}{4M^2 -2Q^2} \, , \nn \\
 && \Omega_L = \frac{\p M}{\p J}\Big|_{Q,\mA_L} = \frac{2a}{\mA_R} = \frac{a}{2M^2 -Q^2} \, , \\
 && \Phi_L = \frac{\p M}{\p Q}\Big|_{J,\mA_L} = \frac{2QM}{\mA_R} = \frac{QM}{2M^2 -Q^2} \, . \nn
\eea
To check that the above expressions hold true, an identity $\mA_+\mA_- = (\mA_R^2 -\mA_L^2)/4 = Q^4 +4J^2$
might be used.

The L-system can represent a rotating charged black hole with mass $M$, charge $Q$, and angular momentum $J$,
its entropy and temperature are given by
\bea\label{ent2}
 S_L &=& \pi\mA_L = \pi(\mA_+ -\mA_-) = S_+ -S_- \nn \\
     &=& 4\pi M\sqrt{M^2 -Q^2 -a^2} \, , \nn \\
 T_L &=& \frac{\kappa_L}{2\pi} = \frac{\sqrt{M^2 -Q^2 -a^2}}{4\pi(2M^2 -Q^2)} \, .
\eea
An important feature of the L-system is that it can reflect the extremality of a KN black hole when it
approaches to its extremal limit, namely, the extremal KN black hole.

It is remarkable to seek the relations between the temperatures of the two newly-constructed systems and
those of the two horizons of the original KN black hole. It is easy to obtain the following relations
\be\label{tem1}
 \frac{1}{\kappa_{R,L}} = \frac{\mA_{L,R}}{\epsilon} = \frac{\mA_+ \mp\mA_-}{\epsilon}
  = \frac{1}{\kappa_+} \pm \frac{1}{\kappa_-} \, ,
\ee
which have an \textit{exact} correspondence in string theory and D-brane physics \cite{GBF}. In particular,
$\kappa_+$ has the same harmonic mean relation as $T_H$ in Eq. (\ref{temp}). In addition, we find that
the angular velocities and electric potentials between the new and old systems are related by
\bea
 && \frac{2}{\Omega_L} = \frac{\mA_R}{a} = \frac{\mA_+ +\mA_-}{a}
  = \frac{1}{\Omega_+} + \frac{1}{\Omega_-} \, , \nn \\
 && \Phi_{R,L} = \frac{\Phi_+ +\Phi_-}{2} +\Big(\frac{\kappa_{R,L}}{\kappa_{L,R}}\Big)
  \frac{\Phi_+ -\Phi_-}{2} \, .
\eea

Note that there is an asymmetry of angular momentum between the R- and L-systems. This fact reflects that
unlike the 5-dimensional case, the addition of angular momentum in four dimension always breaks the right-moving
supersymmetry \cite{GBF}. Without rotation, the D-brane description of an extremal Reissner-Nordstr\"{o}m
black hole can be modelled by a $1+1$-dimensional field theory which turns out to be a $(4, 4)$ superconformal
sigma model. When the angular momentum is included, a black hole can be extremal, but supersymmetry is
broken, leaving us with $(0, 4)$ superconformal symmetry \cite{GBF}. The $N = 4$ superconformal algebra
gives rise to a left-moving $SU(2)$ symmetry of the original $(4, 4)$ chiral superconformal field theory.
The $(0, 4)$ chiral supersymmetry algebra contains an $SU(2)_R$ symmetry, which in fact is the same as
the $SU(2)$ symmetry of global spatial rotations, a symmetry of the Hilbert space of black hole states.
The charge under one $U(1)$ subgroup of this $SU(2)$ group will then be related to the four-dimensional
angular momentum carried by the left movers. The right movers, however, cannot carry macroscopic angular
momentum.

In the Schwarzschild black hole case, the R-system and L-system are indistinguishable, both of them are
identical to its prototype. In other words, the two systems are in thermal equilibrium. It should be
stressed that the above reformulation of the first law of classical black hole thermodynamics are universal
for large classes of black holes in various dimensions. For a Kerr-Sen black hole \cite{Sen} arising
in the low energy effective heterotic string theory, one can obtain similar results with some slight
modifications. [$\mA_{\pm} = 2Mr_{\pm} = 2M(M -b \pm \varepsilon)$, $\Phi_{\pm} = \Phi_{R,L} = Q/(2M) = b/Q$,
where $\varepsilon = \sqrt{(M-b)^2 -a^2}$ in place of $\epsilon$.] Similar formulation can be established
in the case of a three-dimensional BTZ black hole \cite{BTZ} and in other dimensions also.

Next we would like to point out that the above R- and L-systems bear a striking resemblance to the right-
and left-movers in string theory and there exists a close relationship between the thermodynamics of these
systems and that of effective strings or D-branes. Our reformulation of the first law of black hole
thermodynamics can be viewed as an analogy of thermodynamics of effective string or D-brane models.

To make an analogy with string theory more transparent, recall that the thermodynamical object often studied
in string theory and D-brane physics is an effective $1+1$-dimensional ideal gas system composed of a pair of
weakly-coupled oppositely moving states. [In the cases of D-branes, the R- and L-moving massless open string
states on a long brane constitute two noninteracting one-dimensional gas of strings.] 
The effective spatial dimension is one in both cases where the combined system of effective string is divided
into two independent or weak coupling R- and L-sectors. Each sector is a collection of excitation modes and
may be attributed to a gas of strings. The entropy of the combined system is interpreted as a sum of two
independent microscopic contributions from R- and L-moving modes, respectively.

Consider now the thermodynamics of effective strings or D-branes. To this end, we assume that the thermal
equilibrium is maintained in each sector independently, so each sector can be endowed with an effective
temperature. Since massless particles in one spatial dimension are either R-moving or L-moving, all
thermodynamical quantities can be split into a R-moving and a L-moving piece: $E = E_R + E_L$, etc. For
simplicity, we only consider the energy and entropy of two sectors and drop other thermodynamical quantities
(for example, by taking a vanishing chemical potential). It follows that the first laws for the R- and
L-sectors take form
\be
 dE_R = T_R dS_R \, , \qquad dE_L = T_L dS_L \, .
\ee
Taking their sum and difference, we obtain
\be\label{eas}
 dS_{\pm} = dS_R \pm dS_L = \frac{1}{T_{\pm}}dE_+ +\frac{1}{T_{\mp}}dE_- \, ,
\ee
where we introduce $E_{\pm} = E_R \pm E_L$, $S_{\pm} = S_R \pm S_L$, and
\be\label{tem2}
 \frac{2}{T_{\pm}} = \frac{1}{T_R} \pm \frac{1}{T_L} \, .
\ee

We further assume each sector carries energy $E_R = E_L = M/2$, then the total energy is $E = E_+ = M$,
and the total momentum is conserved: $P = E_- = 0$. Thus Eq. (\ref{eas}) reduces to $dE = dE_+ = T_{\pm}
dS_{\pm}$. By comparing it with the formula $dM = \half\kappa_{\pm} d\mA_{\pm} +\cdots$ in Eq. (\ref{FLio})
and observing that Eq. (\ref{tem2}) is essentially the same one as in Eq. (\ref{tem1}), we establish that
\be
 T_{\pm} = \frac{\kappa_{\pm}}{2\pi} \, , \qquad
 S_{\pm} = \pi\mA_{\pm} = \frac{A_{\pm}}{4} \, .
\ee

We find that if the effective strings or D-branes carry the same energy and charges as the black hole, then
their right- and left-moving thermodynamics corresponds \textit{exactly} to the sum and difference of the outer
and inner horizons thermodynamics. This hints that the effective string or D-brane picture of black hole mechanics
may be universal in all cases where the correspondence between black holes and effective strings/D-branes
has been demonstrated. Indeed one may identify the constructed R- and L-systems with the R- and L-movers
in string theory. In this way it is natural to appreciate that black hole entropy can be explained within
string theory and D-brane physics by prescribing a microscopic description for all black holes.

In summary, we have proposed a set of new reformulations of classical black hole thermodynamics, which bears
a close analogy with ``stringy'' thermodynamics. The newly-made systems resemble to the R- and L-movers in
string theory and D-brane physics very much. In our opinion, this resemblance seems to be of a true nature,
so it provides a further evidence to support the effective string/D-brane picture of black holes in the pure
classical thermodynamic side. The formalism presented here suggests a plausible thermodynamic routine to
understand why the mysterious origin problem of black hole entropy could be resolved in the framework of
string theory and D-brane physics. In addition, it may setup a bridge to synthesize various related issues
such as the near-horizon conformal symmetry \cite{SC}, geometric ``conformal field theory'' \cite{CL},
quasinormal and nonquasinormal modes \cite{BC,QNM}, and black hole entropy as well as area spectrum \cite{MVW},
etc. There is clearly much more to be understood.

This project was supported in part by China Postdoctoral Science Foundation and K. C. Wong Education Foundation,
Hong Kong. This work was also supported in part by grants from the Chinese Academy of Sciences and a grant
from the NSF of China under Grant No. 90303002.

\end{document}